# Single femtosecond laser pulse excitation of individual cobalt nanoparticles

Tatiana M. Savchenko[1], Michele Buzzi[1,#], Ludovic Howald[1], Sergiu Ruta[2], Jaianth Vijayakumar[1], Martin Timm[1], David Bracher[1], Susmita Saha[3,4], Peter M. Derlet[5], Armand Béché[6,7], Jo Verbeeck[6,7], Roy W. Chantrell[2], C. A. F. Vaz[1], Frithjof Nolting[1], and Armin Kleibert[*,1]

[1]*Swiss Light Source, Paul Scherrer Institut, 5232 Villigen PSI, Switzerland*

[2]*Department of Physics, University of York, York YO10 5DD, United Kingdom*

[3]*Laboratory for Multiscale Materials Experiments, Paul Scherrer Institute, 5232 Villigen PSI, Switzerland*

[4]*Laboratory for Mesoscopic Systems, Department of Materials, ETH Zurich, 8093 Zurich, Switzerland*

[5]*Condensed Matter Theory Group, NUM, Paul Scherrer Institut, 5232 Villigen PSI, Switzerland*

[6]*EMAT, University of Antwerp, 2020 Antwerpen, Belgium*

[7]*NANOlab Center of Excellence, University of Antwerp, 2020 Antwerpen, Belgium*

*Corresponding author, e-mail: armin.kleibert@psi.ch

**ABSTRACT**

Laser-induced manipulation of magnetism at the nanoscale is a rapidly growing research topic with potential for applications in spintronics. In this work, we address the role of the scattering cross section, thermal effects, and laser fluence on the magnetic, structural, and chemical stability of individual magnetic nanoparticles excited by single femtosecond laser pulses. We find that the energy transfer from the fs laser pulse to the nanoparticles is limited by the Rayleigh scattering cross section, which in combination with the light absorption of the supporting substrate and protective layers determines the increase in the nanoparticle temperature. We investigate individual Co nanoparticles (8 to 20 nm in size) as a prototypical model system, using x-ray photoemission electron microscopy and scanning electron microscopy upon excitation with single femtosecond laser pulses of varying intensity and polarization. In agreement with calculations, we find no deterministic or stochastic reversal of the magnetization in the nanoparticles up to intensities where ultrafast demagnetization or all-optical switching is typically reported in thin films. Instead, at higher fluences, the laser pulse excitation leads to photo-chemical reactions of the nanoparticles with the protective layer, which results in an irreversible change in the magnetic properties. Based on our findings, we discuss the conditions required for achieving laser-induced switching in isolated nanomagnets.

## I. INTRODUCTION

The interaction of light and matter at the nanoscale is a topic of wide interest, with impact on applications ranging from optical manipulation of small objects to bio-imaging and nanoplasmonics [1-3]. In recent years, a number of exciting new light-induced effects have been discovered in magnetism, which promise great potential for applications, for instance, in magnetic data storage, processing or computation [4-10]. Of particular interest in this context is the exploitation of all-optical switching (AOS) effects, where ultrafast laser pulses allow one to control the spin state in individual building blocks in nanoscale magnetic devices [7, 8, 11]. Promising AOS effects, in which femtosecond laser pulse excitation leads to a local reversal of the magnetic moments, have been first observed in ferrimagnetic alloys such as GdFeCo [7, 12, 13]. However, the preparation of such alloys at the nanoscale is difficult due to their high chemical reactivity and most investigations of AOS in these materials are on thin films and on structures with lateral dimensions comparable to the wavelength of the exciting laser pulses [14-16]. More recently, AOS effects have been also observed in ferromagnetic materials such as Co and FePt, for which the preparation of nanomagnets with well-defined properties has been accomplished [17-19]. Hence, ferromagnetic *3d* transition metals and their alloys might serve as prototypical model systems for AOS at the nanoscale. Indeed, AOS has been observed in FePt nanoparticles in granular media, which are used in magnetic data storage [17, 18]. Although the nanoparticles in these media are densely packed and the experiments have probed the average over micrometer-sized regions illuminated by the laser beam, the observations suggest that the AOS in these media occurs through independent magnetic reversals in a large number of individual nanoparticles. Hence, these findings might indicate that AOS can be achieved in isolated nanomagnets, where, in addition to their potential for applications, the intrinsic factors affecting the all-optical switching process may be better characterized. However, the actual magnetic reversal mechanism of AOS in FePt granular media is not yet understood and the conditions needed for achieving AOS in isolated nanomagnets are still largely unexplored.

Presently, two different light matter interaction effects are discussed in the literature, which could lead to a laser pulse-induced spin reversal in nano-sized ferromagnets [18]: (i) thermal heating of the nanoparticle by the laser pulse in combination with magnetic circular dichroism (MCD) giving rise to a stochastic AOS effect resulting in a preferred magnetization direction due to Néel-Brown reversals and (ii) the inverse Faraday effect (IFE), which is essentially a non-thermal, intensity-dependent effect resulting in deterministic AOS by means of angular momentum transfer. In this work, we address experimentally and theoretically, the interaction of ultrashort laser pulses with individual ferromagnetic Co nanoparticles for the observation of AOS effects. We show that, for achieving AOS in isolated nanoparticles, it is necessary to consider laser-induced effects in the nanoparticles and the interaction with the support or the matrix material on an equal footing. Specifically, we find that, because of the inefficient direct energy transfer from the laser pulse to the nanoparticles due to Rayleigh scattering, the interaction of the laser pulse with the surrounding support or matrix material becomes a non-negligible source for heat and hot electrons. The substrate furthermore determines the thermal evolution of the system after the laser pulse is applied and the rate at which Néel-Brown magnetization reversals are possible. Experimentally, we find no evidence for IFE induced AOS in Co nanoparticles for laser pulse intensities below the sample damage threshold, while laser-induced AOS due to MCD and Néel-Brown reversals is suppressed because of the fast heat dissipation into the substrate. Instead, at higher fluences the laser pulse excitation gives rise to photo-chemical reactions of the nanoparticles with the protective layer, which results in an irreversible change in the magnetic properties. The latter effect is not observed in static heating and demonstrates a particular sensitivity of the nanomagnets to femtosecond laser pulse exposure.

The work is organized as follows: In Section II we start by discussing the energy transfer process between the laser pulse and the nanoparticle and its effect on electron, spin, and lattice temperature as predicted by a microscopic three-temperature model. We find that, due to the limited scattering cross section, the number of absorbed photons by the nanoparticle is comparatively small; therefore, only a weak ultrafast demagnetization effect is predicted for laser pulse intensities where significant demagnetization and AOS were observed in densely packed FePt nanoparticle films and below typical sample damage thresholds. Still, the total temperature rise is sufficient to increase the rate for Néel-Brown reversals by several orders of magnitude. In a second step, we include the role of the supporting substrate and the protective layer, Si and C, respectively, in the case considered here. The results show that the laser pulse-induced temperature increase in the C layer actually exceeds that of the Co nanoparticle. Finite element calculations show further that the subsequent heat transfer and cooling of the nanoparticles in the present samples occurs on a time scale that makes switching due to Néel-Brown thermal activation unlikely. These results serve as a background to the experimental results presented in Section III, where a systematic study of the effect of femtosecond laser pulse excitation on magnetism, chemical composition, and morphology of individual, isolated Co nanoparticles using x-ray photoemission electron microscopy (XPEEM) and scanning electron microscopy (SEM) is presented. The data demonstrate that the effect of femtosecond laser pulse excitation of isolated magnetic nanoparticles differs significantly from that of thin films or densely packed nanoparticle ensembles as well as from results obtained by static heating. In Section IV we provide a discussion of the results and propose possible scenarios in which laser-induced deterministic magnetization reversals might be achieved in supported individual nanoparticles.

## II. THEORETICAL CONSIDERATIONS

Femtosecond laser pulse excitation of metallic nanoparticles can lead to a variety of non-thermal and thermal processes ranging from local laser-field enhancement to two-photon excitations or rapid optical heating, which can be exploited for diverse applications in optoacoustic imaging, surface sensing or nanoplasmonics [20-22]. When exciting magnetic nanoparticles with femtosecond laser pulses, the processes that involve the spin system have to be considered in addition to the electronic excitation and the heating of the lattice. In this context, two time scales can be distinguished: (i) the first few picoseconds after the laser pulse, where the primary excited electrons relax and transfer part of their energy to the lattice and to the spin system via electron-electron, electron-lattice, and electron-spin scattering [4]. In these interactions, the temperature of the electron and the spin bath of an isolated nanoparticle can increase by hundreds of K resulting in a partial or full quenching of the magnetization. In the latter case, all information about the initial magnetic state is lost, and the new magnetization direction will be at random, if no additional effects, such as the IFE, imprint a new magnetic state [18]. (ii) After thermalization of electrons, spins, and lattice, the increased temperature can lead to Néel-Brown reversals of the magnetic moment, which in combination with MCD could give rise to a preferred magnetization direction [18, 23]. The occurrence and rate of Néel-Brown reversals depends on the subsequent heat exchange with the surrounding medium/substrate.

In the following we address first the evolution of the electron, spin, and lattice temperature in an isolated nanoparticle using a microscopic three temperature model (3TM) and modeling absorption by Rayleigh's scattering law. In a second step, we consider the heating of the substrate by the laser pulse, the subsequent heat exchange, and its impact on the magnetization of the supported nanoparticle.

**A. Laser pulse induced temperature evolution in isolated magnetic nanoparticles**

The energy $E$ added to an isolated nanoparticle upon excitation with a femtosecond laser pulse is determined by its absorption cross section $\sigma_{abs}$ and the optical peak fluence $\Phi_0$ through the relation [24]:

$$E = \sigma_{abs}\Phi_0 \qquad (1)$$

If the diameter $D$ of a spherical nanoparticle is much smaller than the laser wavelength $\lambda$, the corresponding absorption cross section $\sigma_{abs}$ is given by Rayleigh's scattering law [25]:

$$\sigma_{abs} = 18\pi V(\varepsilon''/\lambda)[(\varepsilon' + 2)^2 + \varepsilon''^2]^{-1} \qquad (2)$$

where $\varepsilon = \varepsilon' + i\varepsilon''$ is the dielectric constant and $V$ the nanoparticle volume. The resulting temperature increase is proportional to the increase of the energy density $w = E/V$ [26]. Accordingly, it follows from (1) and (2), that the temperature evolution in a nanoparticle with $D \ll \lambda$ does not depend on the nanoparticle size, but rather on properties such as $\varepsilon$, $\lambda$, and $\Phi_0$. In a magnetic system, the distinct temperature evolution of electrons, spins, and lattice degree of freedom upon laser pulse excitation can be calculated from $w$ using a phenomenological 3TM [4]. The latter consists of a set of three coupled differential equations, which describe the energy transfer between the three sub-systems with their specific heat capacities $C_p$ and coupling constants $g$ representing the rate of energy exchange between the participating reservoirs. To calculate the corresponding temperature evolution in an isolated nanoparticle, we use the microscopic 3TM as proposed by Koopmans *et al.*, which links the excitation of the spin system to electron–phonon-mediated spin-flip scattering effects and which has been successfully used to describe ultrafast demagnetization in systems such as cobalt [27].

The results of the model calculations are presented in Fig. 1(a) and (b), which show the evolution of the lattice temperature $T_l$, electron temperature $T_e$, along with the relative demagnetization $\Delta m(t) / m(0)$, which reflects the spin temperature $T_s$, upon excitation with a $\tau_l$ = 50 fs (Gaussian FWHM) laser pulse duration, at $\lambda$ = 800 nm wavelength. The calculations were performed for an initial temperature $T$ = 300 K and a peak laser fluence of $\Phi_0$ = 21 mJ cm$^{-2}$. Note that, this fluence corresponds to the highest studied experimentally in this work (see Section III). The dielectric constant $\varepsilon$ is calculated from the index of refraction of Co given in Table 1. All remaining material parameters for Co are taken from Ref. [27] and heat diffusion is omitted to account for an isolated nanoparticle, which is also decoupled from the substrate at this stage of the simulations. Note that the time scale of electronic excitation and subsequent relaxation is too short for the occurrence of thermally induced Néel-Brown magnetization reversals, as discussed below. Hence, the magnetization direction is assumed to be fixed. The calculations show that the electron and lattice temperatures peak at $T_e$ = 500 K and $T_l$ = 420 K, respectively, while the increase in the spin temperature leads to a maximum relative demagnetization $[\Delta m(t) / m(0)]_{max}$ of $-5 \times 10^{-3}$. The latter is significantly lower than that observed in Co thin films, for which experimentally a relative demagnetization of about -0.5 is found at laser peak fluences of about $\Phi_0$ = 5 mJ cm$^{-2}$, but compares well with measurements in Co nanoparticles embedded in $Al_2O_3$ or $SiO_2$ matrices [23, 27]. These results show that the excitation of the spin system in nanoparticles (with $D \ll \lambda$) is significantly less efficient than that of their thin film counterparts as a consequence of the reduced absorption cross section.

As the equilibrium temperature of the nanoparticles in the simulations reaches about 420 K, one would expect to observe an increased frequency of Néel-Brown reversals at longer time scales. To illustrate this, we consider cobalt nanoparticles with a magnetic energy barrier of $E_m$ = 0.63 eV, which are magnetically blocked at room temperature as discussed in Ref. [28].

Using the Arrhenius law for the magnetic switching frequency $\nu = \nu_0 \exp(-E_m/k_B T)$ with an attempt frequency $\nu_0 = 1.9 \times 10^9$ s$^{-1}$ we find that $\nu$ increases by three orders of magnitude from $\nu = 0.05$ s$^{-1}$ at room temperature to 50 s$^{-1}$ to 420 K upon laser pulse excitation. However, whether a thermally activated reversal occurs depends on the time interval at which the particle remains at elevated temperatures as compared to the magnetic switching frequency. This period is determined by the heat exchange with the substrate or matrix material, whose thermal properties need therefore to be considered.

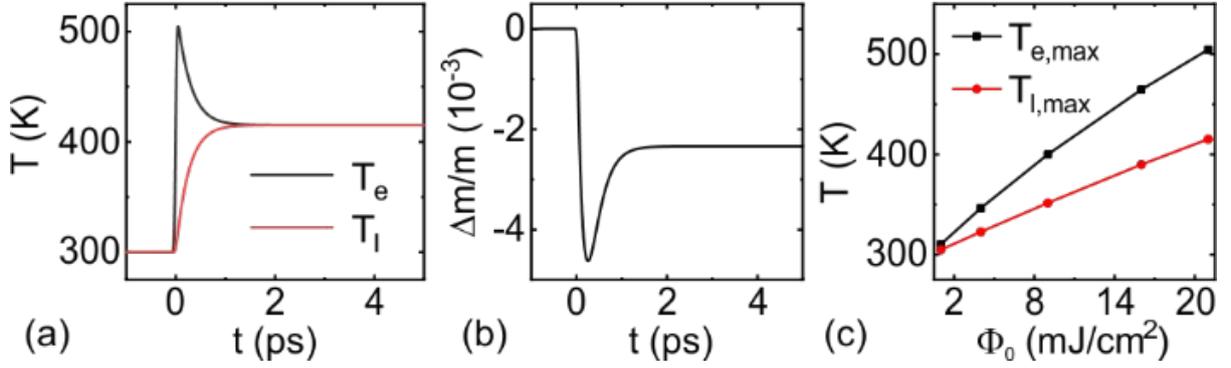

FIG. 1. (a) Calculated time-dependent electron temperature $T_e$, lattice temperature $T_l$, and (b) relative demagnetization $\Delta m/m$ in a thermally isolated cobalt nanoparticle upon excitation with a single femtosecond laser pulse with $\lambda = 800$ nm and a peak fluence $\Phi_0 = 21$ mJ cm$^{-2}$ for $D \ll \lambda$. (c) Maximum electron and lattice temperature as a function of laser fluence.

## B. Temperature profile in the surrounding medium and heat transfer dynamics

Since in practice, the nanoparticle system is supported by a substrate and protected by a capping layer, it is important also to consider the role of light absorption from the surrounding medium. To achieve this, we calculate the temperature profile of the substrate and protection layer after the laser pulse excitation (upon equilibration of electrons and lattice) using a matrix formalism describing light beam absorption and scattering in stratified media [29, 30]. For concreteness, we consider a system consisting of a carbon capping layer, a $SiO_x$ layer, and a semi-infinite Si substrate as used in the experiments. The temperature profile is evaluated by calculating the absorbance given by the local optical constants at the different sections at depth $z$ of the sample and using the respective heat capacities given in Table 1. Taking a grazing angle of incidence 16° for the laser, we obtain the temperature profile shown in Fig. 2(a) for a p-polarized peak laser fluence of $\Phi_0 = 9$ mJ cm$^{-2}$. The calculated temperature profile reveals that the carbon layer is heated to more than 800 K despite its small thickness in comparison to the laser wavelength. In contrast, the $SiO_x$ layer and the Si substrates are much less absorbing and show only a weak temperature increase. For comparison, at this fluence the laser pulse excitation leads to an increase in the nanoparticle temperature to 351 K as shown in Fig. 1(c). Hence, the model calculations suggest that the carbon layer transfers additional heat to the nanoparticles.

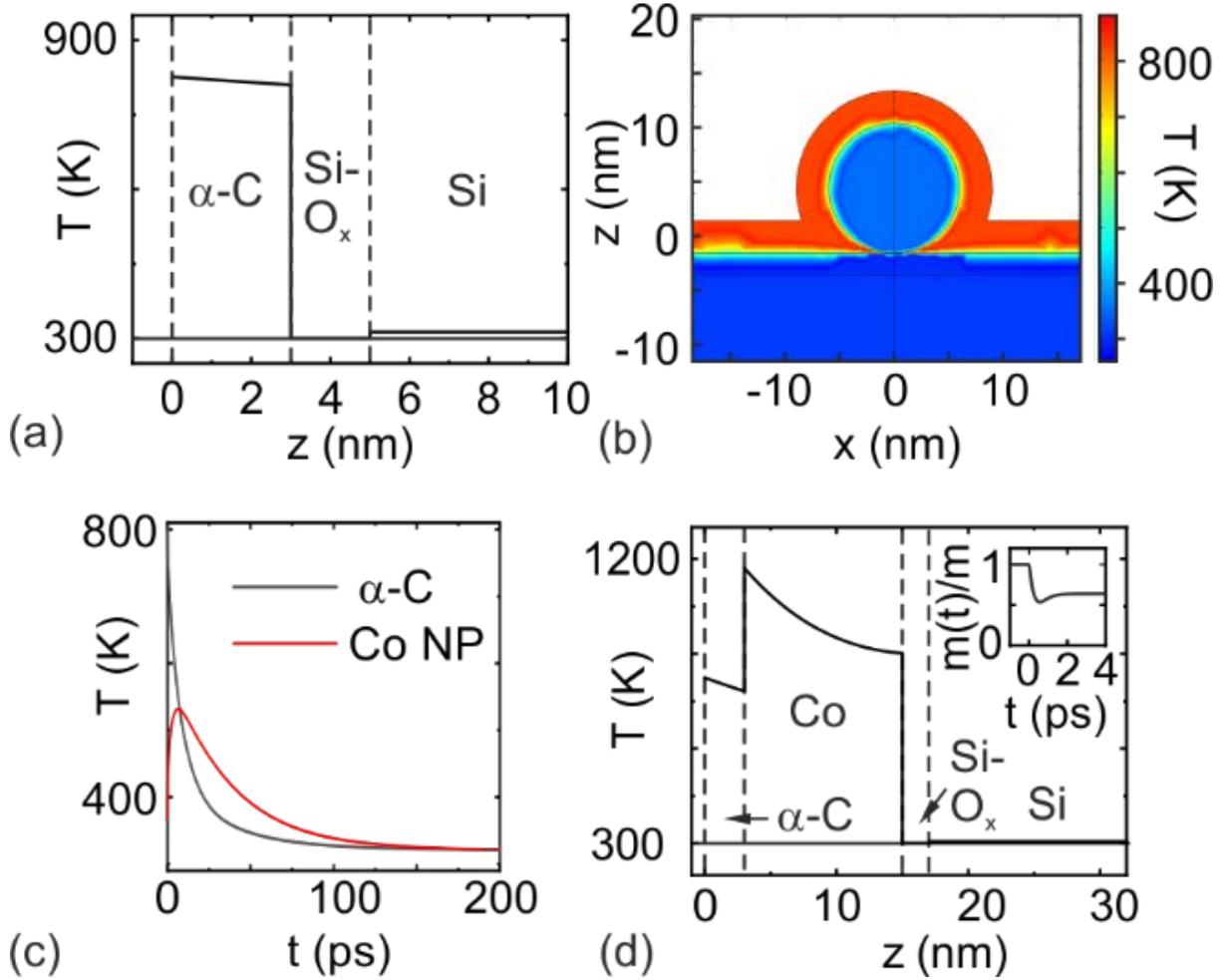

FIG. 2. (a) Calculated instantaneous temperature profile upon absorption of a single pulse with $\Phi_0 = 9$ mJ cm$^{-2}$. (b) Simulated sample geometry and temperature distribution in the sample 1 ps after the laser pulse. (c) Simulated time-dependent temperature evolution in the nanoparticle (red) and in the carbon film (black) due to heat diffusion. (d) Calculated temperature profile in a carbon-capped, 12 nm thick Co film on a Si wafer with a native SiO$_x$ layer with the same laser parameters as in (a). The inset shows the corresponding relative demagnetization calculated using the phenomenological 3TM.

In order to model the heat transfer processes and the related time scales of the supported nanoparticles, including the carbon protection layer, finite element calculations were performed to solve the heat diffusion equation in three-dimensions using COMSOL. The simulated geometry is similar to that introduced before, but now includes a spherical Co nanoparticle with $D = 12$ nm surrounded by the C capping layer as depicted in Fig. 2(b). For initial temperatures we use $T(\alpha\text{-C}) = 826$ K, $T(\text{SiO}_x) = 300$ K, and $T(\text{Si}) = 310$ K as obtained from the calculations shown in Fig. 2(a), while the temperature of the cobalt nanoparticle is set to $T_l(\text{Co}) = 351$ K according to the calculations for a peak laser fluence of $\Phi_0 = 9$ mJ cm$^{-2}$, see Fig. 1(c), and using the thermal conductivity constants $\kappa$ given in Table 1. For simplicity, we ignore the fact that the capping layer on top of the nanoparticle is not a planar film, which might result in a modified light absorption. The simulations further ignore the impact of impedance matching at the nanoscale [31]. The initial temperature distribution in the sample is shown in Fig. 2(b) and the results of the simulation are shown in Fig. 2(c). The simulations reveal that due to the heat transfer from the carbon film to the Co nanoparticle, the latter reaches a peak temperature of $T = 532$ K after about 7 ps. The SiO$_x$ layer is also heated to 390 K within about 5 ps and then cools down. After about 400 ps all components of the sample reach thermal equilibrium in the simulated volume at 321 K. The resulting peak temperature of the Co nanoparticle is indeed significantly higher as compared to direct laser excitation as shown in Fig. 1(c), but still remains far below the Curie temperature $T_c = 1388$ K of bulk fcc cobalt [32, 33]. Hence, significant quenching of the magnetization is not expected. At the peak temperature the magnetic switching frequency of the nanoparticle reaches a value of $\nu = 2000$ s$^{-1}$. The probability that the magnetization of a nanoparticle is switched due to thermal excitation after a time $t$ is given by $P(t) = 1 - \exp(-t\nu)$ [34]. Considering that the temperature after $t = 400$ ps has almost returned to room temperature, this probability is $P = 8.0 \times 10^{-7}$ indicating that laser-induced thermal switching events are unlikely.

Finally, we can compare quantitatively the excitation of the nanoparticles with that of a Co film. In contrast to a spherical nanoparticle, the actual absorption of light by thin films or semi-infinite samples depends on the light polarization, angle of incidence, layer thickness, capping materials, substrates and so forth. In Fig. 2(d) we show for comparison the calculated temperature profile in a carbon-capped, 12 nm thick Co film on a Si substrate with the same photon flux, light polarization, and angle of incidence as for Fig. 2(a). The laser pulse absorption at this flux results in a temperature of about 1200 K, which is much higher than the temperature of 351 K calculated for the nanoparticle. Accordingly, the relative demagnetization of the thin film reaches -0.6 as shown in the inset of Fig. 2(d). This calculation is in good agreement with observations in thin Co films at comparable laser fluences for instance in Ref. 27.

TABLE 1. Parameters used for simulating the initial temperature profile and its temporal evolution: thickness $d$, mass density $\rho$, specific heat capacity $C_p$, thermal conductivity $\kappa$, real ($n$) and imaginary part ($k$) of the index of refraction.

|  | $d$ (nm) | $\rho$ ($10^3$ kg m$^{-3}$) | $C_p$ (J kg$^{-1}$ K$^{-1}$) | $\kappa$ (W m$^{-1}$ K$^{-1}$) | N | k |
|---|---|---|---|---|---|---|
| α-C | 3 | 2.00[a] | 900[b] | 1.0[d] | 2.24[c] | 0.80[c] |
| SiO$_x$ | 2 | 2.21 | 840 | 1.2[e] | 1.45[c] | 0.00[c] |
| Si | ∞ | 2.32 | 712 | 130[f] | 3.69[c] | 0.01[c] |
| Co | - | 8.90 | 420 | 69 – 100 | 2.56[c] | 4.92[c] |

[a] Ref. [35]. [b] Ref. [36]. [c] Ref. [37]. [d] Ref. [38]. [e] Ref. [39]. [f] Ref. [40].

### III. EXPERIMENTAL INVESTIGATION

In this section we show the experimental results of the effect of femtosecond laser pulse excitation on the magnetization of individual cobalt nanoparticles for peak fluences ranging from 1 to 21 mJ cm$^{-2}$ and variable polarization by combining x-ray photoemission electron microscopy (XPEEM) with x-ray magnetic circular dichroism (XMCD) and local x-ray absorption (XA) spectroscopy [41]. When compared to integral methods, this approach allows us to probe unambiguously magnetic reversals or changes in magnetic and chemical properties in individual nanoparticles in large ensembles [42, 43]. The impact of the laser pulses on the morphology of the nanoparticles and the substrate is determined by subsequent scanning electron microscopy (SEM).

**A. Experimental details**

For the XPEEM experiments, Si(001) wafers with a native oxide layer and lithographically prepared gold marker structures are used as substrates. The marker structures serve to identify the very same nanoparticles in XPEEM and SEM. The substrates are introduced into the sample preparation system (base pressure $\leq 5 \times 10^{-10}$ mbar) attached to the XPEEM instrument at the Surface/Interface:Microscopy (SIM) beamline at the Swiss Light Source (SLS) [44]. Upon introduction to the vacuum chamber, the substrates are thermally annealed at 150 – 200 ºC for about 30 minutes to desorb adsorbates originating from ambient air exposure in order to prevent oxidation of the cobalt nanoparticles upon contact with the substrate. An ultrahigh vacuum (UHV) compatible arc cluster ion source (ACIS) is used to deposit mass-filtered pure, metallic cobalt nanoparticles with diameters $D$ varying from 8 to 20 nm on the substrates held at room temperature [45]. The deposition occurs under so-called soft landing conditions, where the kinetic energy of the nanoparticles is below 0.1 eV atom$^{-1}$, low enough to prevent fragmentation or damage to the substrate upon landing [45-47]. Using a gold mesh as a flux monitor in the beam of the electrically charged nanoparticles, the nanoparticle density on the substrate is set to about one nanoparticle per µm$^2$ to avoid inter-particle interactions and to allow us to resolve individual nanoparticles in XPEEM at a spatial resolution of about 50 nm. Finally, the sample is covered with 2 – 3 nm of amorphous carbon to prevent chemical reactions of the nanoparticles with residual gas molecules during the experiments, which typically involve hours of exposure to intense x-ray and laser radiation [5]. Carbon films of this thickness are electrically conductive and transparent to electrons and x-rays, and are therefore ideally suited to XPEEM investigations. A schematic diagram of the sample and of the experimental setup is shown in Fig. 3(a). After nanoparticle and carbon deposition the samples are transferred under UHV conditions to the XPEEM instrument, which has a base pressure $< 5 \times 10^{-10}$ mbar [44]. In a second experiment, samples with an additional carbon layer between the silicon substrates and the nanoparticles, to avoid a direct contact of the nanoparticles with the native Si oxide layer, were investigated. Finally, a reference sample for high resolution scanning transmission electron microscopy (HR-STEM) investigations was prepared by depositing cobalt nanoparticles under similar conditions directly on a 10 nm $Si_3N_4$ membrane (TEMwindows.com) and capped with C to prevent oxidation during transfer to the HR-STEM instrument. Prior to the nanoparticle deposition the $Si_3N_4$ membranes were annealed for about 30 minutes at about 150 – 200 ºC to remove adsorbates. To avoid damage of the $Si_3N_4$ membrane the temperature is increased in small steps over a course of 30 minutes. After the nanoparticle deposition, which is carried out after cooling back to room temperature, the sample is capped with an amorphous carbon layer.

In XPEEM, the sample is illuminated with x-rays at a grazing angle of incidence $\theta_k = 16°$ as shown in Fig. 3(a). The position of the deposited Co nanoparticles on the substrate is visualized by XPEEM images obtained by pixelwise division of images recorded with the photon energy set to the Co $L_3$ edge (781 eV) and a so-called flat-field image for which the microscope is defocused. This process removes inhomogeneity artifacts of the detector from the data. At this photon energy, the cobalt nanoparticles are resonantly excited and appear as bright spots on the darker substrate background. Magnetic characterization of the nanoparticles is achieved using the x-ray magnetic circular dichroism (XMCD) effect at the Co $L_3$ edge [48]. The XMCD effect gives rise to a magnetization dependent intensity according to $I(C^{\pm}) = I_0 \pm \gamma (\vec{k} \cdot \vec{m})$, where $I_0$ is the isotropic (non-magnetic) contribution, $\vec{k}$ is the x-ray propagation vector, $\vec{m}$ is the magnetization vector of the particle, $\gamma$ is a material and photon energy dependent constant, and $C^{\pm}$ denote circular right/left-handed polarization. Magnetic contrast maps are obtained by pixel-wise calculation of the asymmetry, $[I(C^+) - I(C^-)]/[I(C^+) + I(C^-)]$, of two images recorded with $C^+$ and $C^-$ polarization, respectively, with the photon energy set to the Co $L_3$ edge using the tune-detune mode at the SIM beamline [44,49]. The resulting magnetic contrast of individual nanoparticles is proportional to $\vec{k} \cdot \vec{m}$, and, hence, can range from black to white depending on the actual orientation of $\vec{m}$ with respect to $\vec{k}$. In addition, only nanoparticles with a magnetic relaxation time $\tau_m$ larger than or equal to the experimental time $\tau_x = 400$ s required to acquire magnetic contrast maps can exhibit magnetic contrast. Chemical characterization of the nanoparticles is achieved through local XA spectroscopy by recording image sequences with linearly polarized x-rays across the Co $L_3$ edge. XA spectra are obtained by extracting the image intensities from small areas, typically $5 \times 5$ pixels, centered on the position of individual nanoparticles in the XPEEM images. These data are normalized to the background signal, extracted from an area of the same size next to the nanoparticle [41]. For the present work, spectra of about 30 nanoparticles have been averaged. All raw XPEEM image sequences are first flat field- and drift-corrected before further data analysis is carried out.

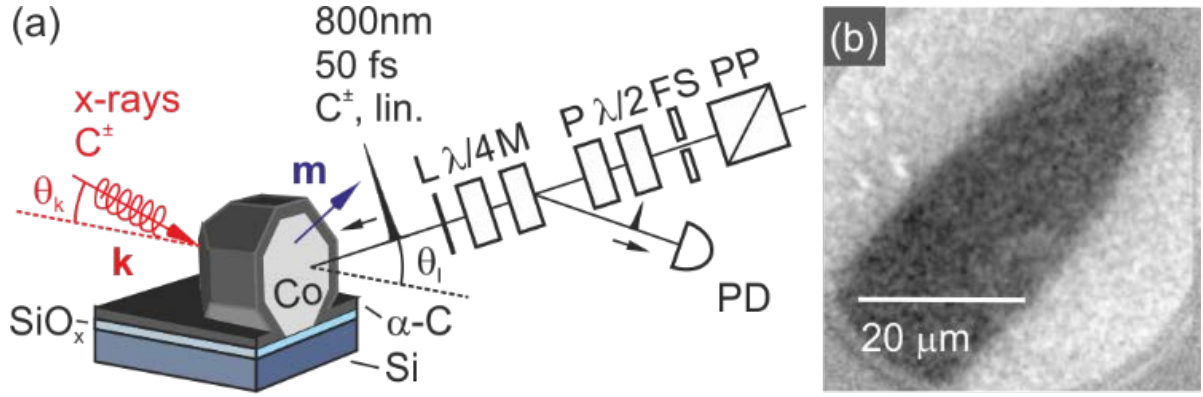

FIG. 3. (a) Schematic diagram of the sample and the experimental set-up showing the direction of the incoming x-ray and laser light and the set-up for generation of single fs laser pulses. PP: pulse picker, FS: fast shutter, $\lambda/2$, $\lambda/4$: wave plates, P: polarizer, M: leaking mirror, P: polariser, L: lens, and PD: fast photodiode. (b) Laser beam profile (dark area) determined using a Cs-covered reference sample in the XPEEM instrument.

A Ti:sapphire oscillator (XL-500, Femtolasers GmbH) with a wavelength $\lambda = 800$ nm, a pulse energy $E = 500$ nJ, a pulse duration $\tau = 50$ fs, and a repetition rate of 5.2 MHz is used to excite the sample. The laser beam is aligned in the XPEEM instrument using a Cs covered sample, which permits direct imaging of the laser spot as shown in Fig. 3(b). A schematic diagram of the laser set-up is given in Fig. 3(a). The laser beam impinges the sample at a grazing angle $\theta_l = 16°$ through a strain-free fused silica UHV viewport in order to avoid modifications of the laser beam polarization by strain-induced birefringence [50]. The laser pulse energy at the sample is set using a half wave plate and a polarizing beam splitter. A fast photodiode is used to monitor the intensity of each laser pulse in a reference beam during the experiments. The intensity at the sample is initially calibrated using a photodiode mounted to a sample holder and measuring the laser intensity for given settings at the sample position in the XPEEM instrument. All pulse energies are given relative to this measurement. The grazing incidence gives rise to an elliptical laser spot profile with dimensions of $FWHM_x = (20 \pm 5)$ µm and FWHM$_y$ = (73 ± 18) μm. Assuming an elliptical Gaussian intensity distribution the peak fluence can be calculated as $\Phi_0$ = 4ln(2)$E$/(πFWHM$_x$ × FWHM$_y$). Table 2 shows the investigated laser pulse energies and the resulting peak fluences, together with the corresponding photon densities at the sample. A quarter wave plate is used to switch the polarization of the laser pulses between linear, $C^+$ and $C^-$. Single laser pulses are selected using a pulse-picker (PP) and a fast mechanical shutter (FS) [15]. All experiments are performed at room temperature. The morphology of the sample and the nanoparticles are investigated by means of SEM and high resolution scanning transmission electron microscopy (HR-STEM), respectively. The atomic resolution HR-STEM investigations with high-angle annular dark-field (HAADF) imaging are carried out using a FEI Titan³ equipped with a Cs probe corrector. The microscope is operating at 300 kV with a convergence angle of 20 mrad allowing a maximum spatial resolution of 70 pm.

TABLE 2. Investigated laser pulse energies, peak fluences, and respective photon densities on the sample with estimated errors.

| Laser pulse energy $E$ (nJ) | 14 ±1 | 65 ± 3 | 150 ± 8 | 270 ± 14 | 352 ± 18 |
|---|---|---|---|---|---|
| Peak fluence $\Phi_0$ (mJ cm$^{-2}$) | 1 ± 1 | 4 ± 1 | 9 ± 3 | 16 ± 6 | 21 ± 8 |
| Photon density $n_{Ph}$ (nm$^{-2}$) | 34 ± 10 | 160 ± 60 | 370 ± 130 | 660 ± 240 | 860 ± 310 |

**B. Experimental results**

Figure 4(a) shows an XPEEM image of the sample. The displayed region represents only a small area of the investigated field of view, which centered on the laser spot. Figure 4(b) shows the corresponding magnetic contrast map. In accordance with our previous work on similar cobalt nanoparticle samples, but with no C capping, a large portion of nanoparticles (approximately 50%) exhibits stable magnetic contrast ranging from white to black [28]. This contrast distribution reflects a random orientation of the magnetization $\vec{m}$ of magnetically blocked nanoparticles with respect to the x-ray propagation vector $\vec{k}$ due to the stochastic nature of the deposition process [42]. Several magnetically blocked nanoparticles with $\tau_m > \tau_x$ are highlighted with solid circles in Fig. 4(b). Similar to the earlier experiments we also find a number of particles that exhibit no magnetic contrast, for instance those, highlighted with dashed circles in Figs. 2(a) and (b). As discussed in Ref. [28] the absence of magnetic contrast can be either assigned to superparamagnetic states with $\tau_m < \tau_x$ or to magnetically blocked nanoparticles with $\vec{m} \perp \vec{k}$. In nanoparticles with $\tau_m \approx \tau_x$, spontaneous, thermally induced magnetization reversals can be directly observed as a function of time in consecutively recorded magnetic contrast maps as shown further below. Further, it was shown in Ref. [28] that magnetically blocked and superparamagnetic states occur irrespective of the size of the (near spherical) nanoparticles. Comparing the pristine state of the sample in Figs. 4(a) and (b) and the results in Ref. [28] suggests, that the carbon capping has no or only a minor impact on the magnetic behavior of the cobalt nanoparticles. The XA spectrum shown in Fig. 4(c) further confirms the metallic state of the nanoparticles prior to the laser-based experiments. Finally, the inset of Fig. 4(b) displays an HAADF HR-STEM image of a near spherical Co nanoparticle with a carbon-capping.

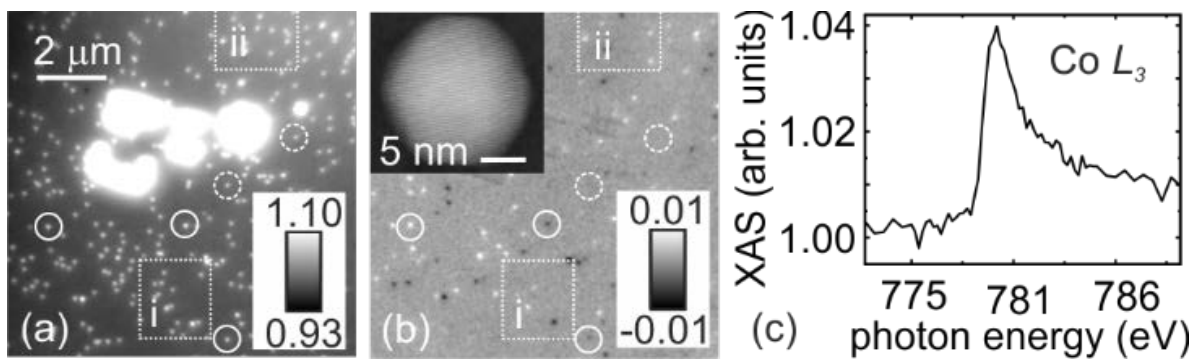

FIG. 4. (a) XPEEM image recorded with the photon energy set to the Co $L_3$ edge. Bright spots correspond to cobalt nanoparticles. Marker structures appear as saturated bright features. (b) Magnetic contrast map of the same sample area. Solid circles highlight magnetically blocked nanoparticles. Dashed circles highlight nanoparticles without magnetic contrast. The inset shows a HAADF STEM image of a cobalt nanoparticle capped with an amorphous carbon layer. (c) XA spectra of the nanoparticles recorded at the Co $L_3$ edge prior to the laser experiment. The grey tone intensities in (a) and (b) are set to visualize the nanoparticle signals.

In order to distinguish spontaneous magnetization reversals in thermally active nanoparticles from laser-induced effects on the magnetization, a control sequence of XPEEM images and magnetic contrast maps without laser excitation are recorded. The upper row "A" in Fig. 5 displays ten XPEEM images "a – j" of the control sequence in area "i" of the sample, which is denoted by the respective dashed box in Figs. 4(a) and (b). For the sake of brevity, only a representative, small region of the sample containing about 20 nanoparticles out of the 500 present in the laser illuminated area is highlighted. The respective magnetic contrast maps, shown in row "B", reveal a number of magnetically blocked nanoparticles with stable magnetic contrast throughout the entire sequence from "a – j" such as for instance nanoparticle "1", which is highlighted with a solid circle. Similarly, a number of nanoparticles with no magnetic contrast or varying contrast are found. For instance, nanoparticle "2" exhibits no contrast throughout the sequence, while nanoparticle "3" is thermally active exhibiting white magnetic contrast in "a – f" and black magnetic contrast in "i" and "j". Row "C" displays a sequence of magnetic contrast maps upon consecutive excitation with single linearly polarized laser pulses with a pulse energy of $E = 65$ nJ ($\Phi_0 = 4$ mJ cm$^{-2}$). While single pulse AOS has been found at similar fluence in Co thin films [51], we find that none of the magnetically blocked nanoparticles in Row "C" is affected by the laser pulse excitation, see for instance nanoparticle "1". In fact, in the full dataset of 500 nanoparticles, only one magnetically blocked nanoparticle changed its magnetic contrast once upon excitation with the laser pulse, see nanoparticle "3". Also, most of the nanoparticles found without magnetic contrast in the control sequence, such as nanoparticle "2" in "B", remain without magnetic contrast upon laser excitation. In two cases in the investigated area, the onset of magnetic contrast is observed upon laser pulse excitation, which could be due to a laser-induced increase of the magnetic energy barrier or due to a spontaneous modification in the magnetization direction. Similar results are obtained when exciting the sample with $C^+$ and $C^-$ laser polarization without evidence for reversals induced by IFE or MCD in individual nanoparticles at this fluence (not shown).

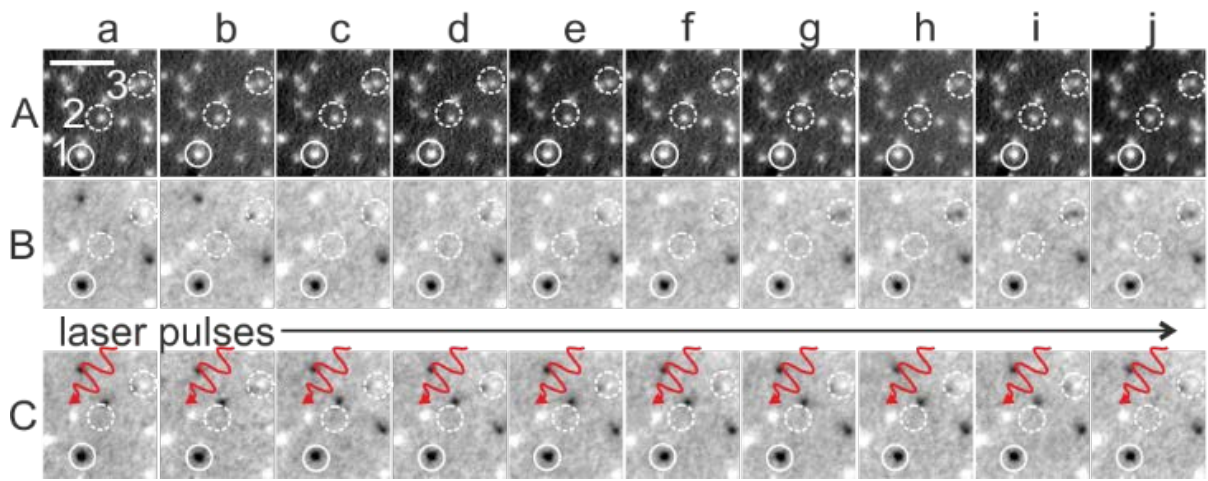

FIG. 5. Row "A": XPEEM images of region "ii" in Fig. 4. Row "B": Control sequence of ten consecutively recorded magnetic contrast maps without laser exposure. Row "C": Similar sequence, but each magnetic contrast map is recorded upon exposure to a single laser pulse with a peak fluence of $\Phi_0 = 4$ mJ cm$^{-2}$ and linear polarization. The scale bar in panel (a) in row "A" corresponds to 1 μm. The solid circle denotes a magnetically blocked nanoparticle, while the two dashed circles highlight nanoparticles, which exhibit no or varying contrast over the time of the experiment. The grey tones range from 0.97 (black) to 1.06 (white) in row A and from -0.07 (black) to +0.07 (white) in rows B and C.

Similar results are found upon excitation at increasingly higher laser pulse fluence. In this excitation regime, a significantly larger number of nanoparticles manifested an irreversible loss of magnetic contrast during the course of the experiments. This behavior became dominant for laser pulses at $\Phi_0 = 9$ mJ cm$^{-2}$ and higher, where a sizeable proportion of the nanoparticles lost their magnetic contrast with each laser pulse. This is shown in Fig. 6 for laser pulses with linear polarization and $\Phi_0 = 9$ mJ cm$^{-2}$. Similar to Fig. 5, rows "A" and "B" show first the data of the control sequence for region "ii" of the sample as shown by the dashed box in Figs. 4(a) and (b). Three particles "1 – 3" are highlighted. Nanoparticle "1" exhibits stable (black) magnetic contrast in the control sequence, which is also not affected by the laser pulses as seen in row "C". Nanoparticle "2" shows no magnetic contrast throughout both the control and the laser exposure series. Nanoparticle "3" shows varying magnetic contrast in the control sequence and further contrast reversals are observed after the first six laser pulses, see panels "a – f" in row "C" of Fig. 4. However, after the sixth laser pulse the magnetic contrast of the nanoparticle does not reappear. A similar loss of magnetic contrast is found in most magnetically blocked or thermally active nanoparticles in row "C" and in the rest of the sample, cf. panels "a" and "j" in row "C". For higher laser pulse energies, the number of nanoparticles with an irreversible loss of magnetic contrast increases per pulse. In the few nanoparticles, which maintain their magnetic contrast, even higher laser pulse energies still resulted in no detected laser-induced magnetic reversal, irrespective of the laser pulse polarization.

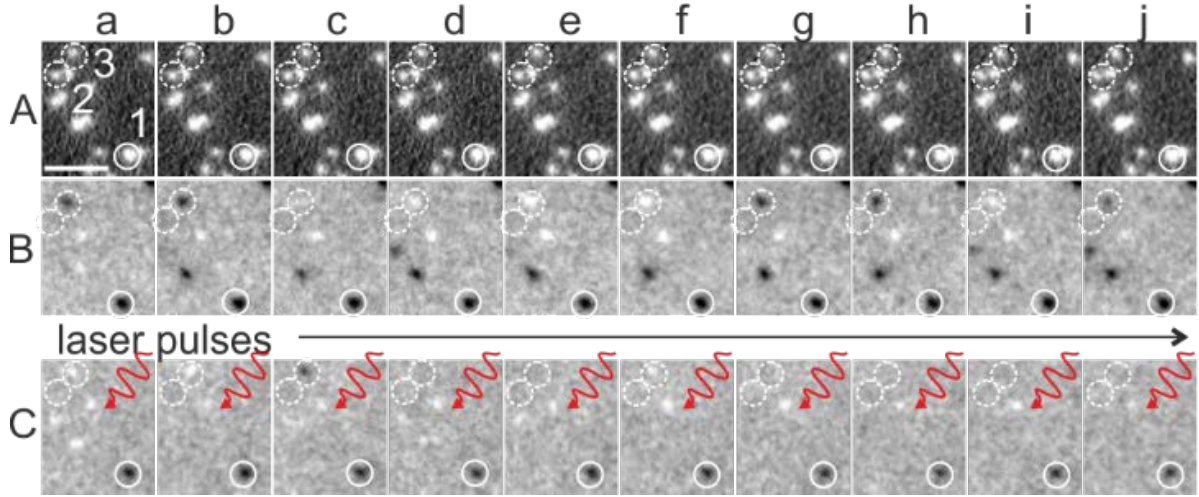

FIG. 6. Row "A": XPEEM images of region "ii" in Fig. 4. Row "B": Control sequence of ten consecutively recorded magnetic contrast maps without laser exposure. Row "C": Similar sequence, but each magnetic contrast map is recorded upon exposure to a single laser pulse with $\Phi_0 = 9$ mJ cm$^{-2}$ and linear polarization. The scale bar in panel "a" in row "A" corresponds to 1 µm. The solid circle denotes a magnetically blocked nanoparticle, while the two dashed circles highlight nanoparticles, which exhibit no, or varying contrast over the time of the experiment. The grey tones range from 0.97 (black) to 1.06 (white) in row A and from -0.07 (black) to +0.07 (white) in rows B and C.

XA spectra recorded after the full series of laser exposure experiments with fluences up to $\Phi_0$ = 21 mJ cm$^{-2}$ reveal a shoulder at 782 eV next to the metallic cobalt peak, see Fig. 7(a), which indicates a change in the chemical state. Since the nanoparticles are in contact with C and SiO$_x$, the new peak could indicate for instance the formation of Co oxide or a Co-C compound. A comparison with reference spectra of fcc CoO and Co$_3$O$_4$ in Fig. 7(b) and with spectra of a mixed C-Co phase in Fig. 7(c) suggests a carbide formation. In a second series of experiments, the Si wafer was capped with an additional carbon layer before the nanoparticle deposition, so that the nanoparticles were fully embedded in carbon. These experiments gave the same result as above, with XA spectra similar to that of Fig. 7(a), which corroborates the laser-induced formation of mixed Co-C phases. The loss of magnetic contrast in the chemically modified nanoparticles is assigned to a reduction of the magnetic volume and of the magnetic energy barrier. We note that for single crystalline $Co_3C$ nanoparticles ferromagnetic order and surprisingly high magnetic energy barriers have been reported [52]. However, the loss of magnetic contrast in the present experiments suggests that no such phase is formed. SEM images of the laser-exposed areas show no hint of macroscopic sample damage such as laser ablation, which in $SiO_x$ occurs at much higher fluences [53]; see the dashed area in Fig. 7(d). High resolution SEM images of individual, laser-exposed nanoparticles are shown in Figs. 7(e) and (f). These particles appear somewhat larger with some irregularities when compared to nanoparticles of the same sample that were not exposed to the laser, see Figs. 7(g) and (h), suggesting a change in morphology due to the chemical reaction.

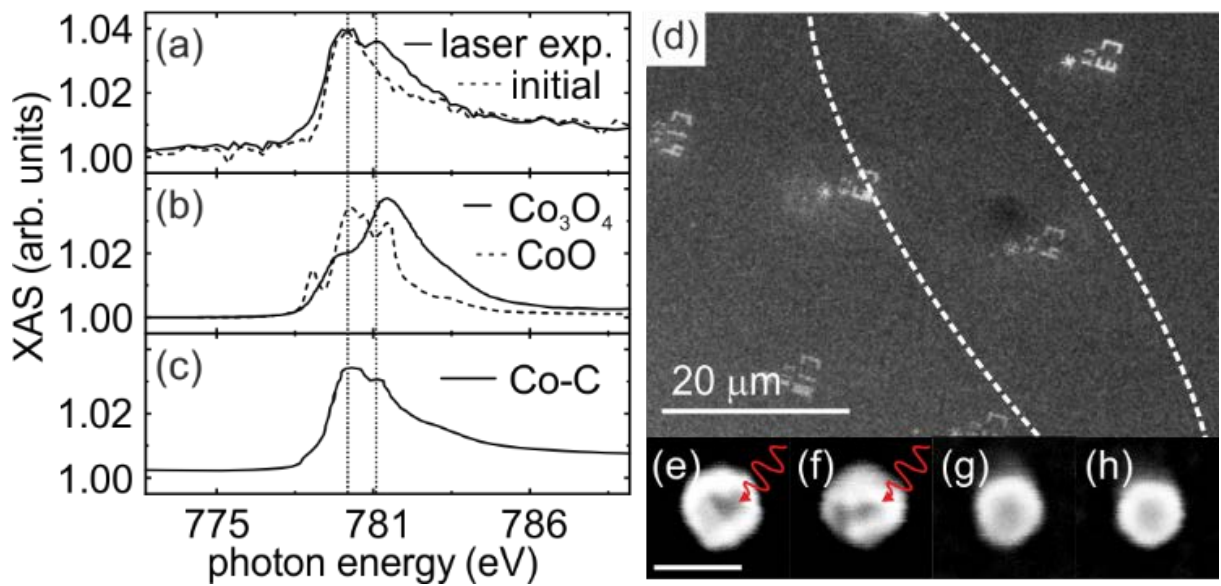

FIG. 7. (a) XA spectra of the nanoparticles before (dashed line) and after laser excitation with $\Phi_0 = 9$ mJ cm$^{-2}$ (solid line). (b) XA reference spectra for $Co_3O_4$ and fcc CoO from Refs. [54, 55]. (c) XA reference spectra of cobalt nanoparticles with a size of about 3 nm being embedded in a carbon matrix from Ref. [56]. The dashed lines denote the peak positions in the spectra

recorded after the laser excitation experiments. (d) SEM image of the laser exposed area of the sample recorded after the laser experiments. The dashed line denotes the position of the laser spot on the sample. (e, f) SEM images of two laser-exposed cobalt nanoparticles. (g, h) SEM images of two cobalt nanoparticles from the same sample, but from a region that was not exposed to the laser. The scale bar corresponds to 10 nm.

## IV. DISCUSSION

Our experiments show that neither deterministic IFE nor stochastic MCD induced magnetic reversals occur in isolated Co nanoparticles upon femtosecond laser pulse excitation with peak fluences up to 21 mJ cm$^{-2}$. For the IFE, this reveals that up to these intensities the laser pulse induced magnetic moment in isolated Co nanoparticles is too small to reverse their magnetization as discussed for the case of FePt nanoparticles in Ref. [18]. The IFE induced change in the magnetization is given by $\Delta M = K_{IFE}\, I/c$ with $K_{IFE}$ being the inverse Faraday constant, $I$ the laser pulse intensity, and $c$ the velocity of light [18, 57]. Assuming that, for non-thermal switching, the induced magnetic moment $\Delta M$ in a Co nanoparticle has to be comparable to its saturation magnetization and considering that we observe no switching for fluences up to 21 mJ cm$^{-2}$, we can estimate an upper limit for $K_{IFE}$. Taking the reduced absorption due to Rayleigh scattering into account, we find $|K_{IFE}| < 10$ T$^{-1}$. This limit is consistent with the much smaller value of $K_{IFE} = 0.025$ T$^{-1}$ being predicted by *ab initio* calculations for Co at $\lambda = 800$ nm [57]. In Ref. [18] it was proposed that a magnetic reversal by means of the IFE might be possible, if the laser excitation simultaneously leads to a significant quenching of its magnetization so that the IFE induced magnetic moment becomes the dominant contribution. However, our calculations show that for isolated Co nanoparticles the laser-induced demagnetization is small due to the low absorption cross section and the comparably high Curie temperature of Co, in agreement with other experiments [23]. Hence, thermally assisted IFE in isolated Co nanoparticles will be difficult to achieve. Although heat exchange with the surrounding medium can lead to an additional increase in the nanoparticle temperature, it remains unclear whether the temporal evolution of the temperature is compatible with the timing requirements for the proposed heat-assisted IFE [18]. The absence of laser-induced Néel-Brown reversals and the related stochastic AOS due to MCD in the present experiments can be assigned to the fast heat transfer to the Si substrate and the rapid cooling of the nanoparticles as shown in Section II. This property can be utilized to unambiguously discriminate stochastic MCD from deterministic IFE induced AOS in nanoparticles. Experimental evidence for a substrate effect on the efficiency of laser pulse induced magnetic switching in granular FePt nanoparticle media was reported in Ref. [58], where a free standing layer of FePt nanoparticles showed a larger number of magnetic switching events compared to FePt nanoparticles on a layered substrate optimized for heat-assisted magnetic recording. This may suggest that MCD and Néel-Brown reversals also play a role in the laser-induced magnetic switching observed in Ref. [18]. Our simulations show that the heat exchange between nanoparticles and the matrix or medium and its dynamics need to be considered in order to evaluate the impact of thermal effects on the magnetization of the nanoparticles such as Néel-Brown reversals. Note that the actual heat transport at the nanoscale might differ from the simulations discussed above due to impedance matching. The latter could lead to slower heat transport and longer dissipation times when compared to the data in Fig. 2(c) [31]. Hence, for a more accurate description and design of nanostructures for AOS this aspect requires additional consideration.

Our investigations reveal further, that the interaction of femtosecond laser pulses with isolated nanoparticles leads to effects, which have no counterpart in static experiments and, hence, require particular attention when addressing AOS in magnetic nanoparticles. One such effect concerns the observed laser-induced formation of a Co-C phase at higher laser pulse intensities. Co-C phases have been found for instance in cobalt nanoparticles with sizes of about 3 nm embedded in a carbon matrix [56]. In static experiments, it was found that annealing to 750 K leads to a decomposition of the mixed Co-C phase and the formation of fcc Co and graphitic-like carbon [59, 60]. However, the present experiments show that the opposite reaction can occur upon femtosecond laser pulse exposure at peak fluences of $\Phi_0 = 9$ mJ cm$^{-2}$ or higher, where the C-layer reaches temporarily a temperature of 800 K according to our simulations. In the present experiments, we assign the formation of the Co-C phase to a photo-chemical reaction triggered by hot electrons created by the laser pulse [61]. The stability of this phase could be due to the short period at which the sample is at elevated temperature. We may note that an irreversible sample modification upon femtosecond laser pulse exposure was also noticed in the case of the granular media consisting of FePt nanoparticles, and was assigned to a laser-induced damage of the C-matrix, showing that such effects are not specific to the Co-C system [18]. Another effect concerns the stability of the magnetic energy barriers of the investigated Co nanoparticles. Previous experiments on similar Co nanoparticles revealed that static heating to 470 K caused an increase of the magnetic energy barrier in a large number of particles, which lead to irreversible transitions from a superparamagnetic to a magnetically blocked state in XPEEM investigations [28]. In contrast, in the present experiments, only very few nanoparticles exhibited a comparable transition to a magnetically blocked state upon laser pulse exposure, although our simulations show that the laser pulse excitation at $\Phi_0 = 9$ mJ cm$^{-2}$ results in a peak temperature of $T = 532$ K in the Co nanoparticles. These findings illustrate that some important effects from femtosecond laser pulse exposure, including the induced temperature spikes, are not predicted by extrapolation from static heating experiments.

Finally, we discuss the absence of laser-induced magnetization reversals in the investigated isolated Co nanoparticles, in contrast to the observation of AOS in the granular FePt nanoparticle media, which highlights an important difference between fs laser pulse excitation of isolated nanoparticles and that of dense ensembles of nanoparticles, related to optical coupling effects. Nanoparticles with distances smaller than 4 or 5 times their diameter are optically coupled when excited by a laser pulse, which leads to near-field effects and increased absorption [24]. The arrangement of the nanoparticles determines further the subsequent heat diffusion processes, including the temporal and spatial temperature evolution in the substrates and matrices [24]. These effects impact AOS phenomena in dense nanoparticle arrays and can amplify both the IFE but also MCD and Néel-Brown reversals and complicate the discrimination of non-thermal and thermal effects. A first evidence for the impact of near-field effects have been reported in case of the FePt nanoparticles in granular media [58]. However, while this effect might be used to achieve AOS in smaller isolated clusters of FePt nanoparticles, it prevents the optical control of individual nanomagnets. A solution to this issue might be to combine isolated nanomagnets with non-magnetic plasmonic nanoantennas to locally enhance the electric laser field [9, 10, 62]. Such approaches are currently under intense investigation, but the preparation of nanostructures with the desired magnetic and optical properties remains challenging. The present experiments and the discussions in the literature suggest that one key issue for achieving deterministic AOS in isolated nanoparticles is related to quenching the magnetization at laser pulse intensities below the sample damage threshold. Hence, nanoparticle systems with lower Curie temperatures than that of FePt with $T_c$ = 750 K or Co with $T_c$ = 1388 K could be worth investigating [31, 32, 63]. A lower $T_c$ could be achieved for instance in nanoparticles of 3$d$ transition metal alloys or by nanoparticle size reduction [64]. Alternatively, the substrate and matrix materials could be optimized to further increase the nanoparticle temperature. Finally, our simulations suggest that the critical temperatures of the prototypical ferrimagnetic AOS material GdFeCo, with $T_c$ = 500 K and $T_{comp}$ = 450 K, could be achieved in isolated nanoparticles [7, 65]. However, the preparation of nanoparticles of this alloy has yet to be achieved.

## V. CONCLUSIONS

We have investigated the effect of single femtosecond laser pulses on the magnetic and chemical state of individual cobalt nanoparticles with sizes ranging from 8 to 20 nm, as a function of laser pulse energy and polarization. The experiments revealed no evidence for laser-induced all-optical switching of the magnetization in the investigated cobalt nanoparticles, irrespective of laser polarization or laser intensity. The investigated fluences covered ranges where demagnetization or all-optical switching is commonly observed in thin film samples. Above fluencies of about 9 mJ cm$^{-2}$ a laser-induced chemical reaction of the nanoparticles with the carbon capping layers was observed, but no indication for other damage such as laser ablation was found. Calculations based on a microscopic three-temperature model, local absorbance of the laser pulse, and finite element methods show that the absence of magnetic switching can be assigned to the reduced direct laser absorption of the nanoparticles due to Rayleigh scattering and the comparably high Curie temperature of cobalt. The rapid heat dissipation after the laser pulse excitation prevents further laser-induced thermal fluctuations of the magnetization. Finally, we propose possible pathways to achieve all-optical switching in isolated nanomagnets based on the present findings.

**ACKNOWLEDGEMENTS**

This work received funding by the Swiss National Foundation (SNF) (Grants. No 200021_160186 and 2002_153540), the Swiss Nanoscience Institute (SNI) (Grant. No SNI P1502), the European Union's Horizon 2020 research and innovation programme under Grant Agreement No. 737093 (FEMTOTERABYTE), and the COST Action CA17123 (MAGNETOFON). Part of this work was performed at the Surface/Interface: Microscopy (SIM) beamline of the Swiss Light Source (SLS), Paul Scherrer Institut, Villigen, Switzerland. Part of the simulations were undertaken on the VIKING cluster, which is a high-performance compute facility provided by the University of York. We kindly acknowledge Anja Weber from PSI for preparation of substrates with marker structures. A.B. and J.V. (Antwerp) acknowledge funding through FWO project G093417N ('Compressed sensing enabling low dose imaging in transmission electron microscopy') from the Flanders Research Fund. J.V. (Antwerp) acknowledges funding from the European Union's Horizon 2020 research and innovation program under grant agreement No 823717 – ESTEEM3. S.S. acknowledges ETH Zurich Post-Doctoral fellowship and Marie Curie actions for people COFUND program.

[#]Present address: Max Planck Institute for Structure and Dynamics of Matter, Hamburg, Germany.

**REFERENCES**

1. P. Biagioni, J.S. Huang, and B. Hecht, "*Nanoantennas for Visible and Infrared Radiation*", Rep. Prog. Phys. **75**, 024402 (2012).

2. M. Dienerowitz, M. Mazilu, and K. Dholakia, "*Optical Manipulation of Nanoparticles: a Review*", J. Nanophotonics **2**, 021875 (2008).

3. L. Cognet, S. Berciaud, D. Lasne, and B. Lounis, *"Photothermal Methods for Single Nonluminescent Nano-Objects"*, Anal. Chem. **80**, 2288 (2008).

4. E. Beaurepaire, J. C. Merle, A. Daunois, and J. Y. Bigot, *"Ultrafast Spin Dynamics in Ferromagnetic Nickel"*, Phys. Rev. Lett. **76**, 4250 (1996).

5. A. V. Kimel and M. Li, *"Writing Magnetic Memory with Ultrashort Light Pulses"*, Nat. Rev. Mater. **4**, 189 (2019).

6. M. Pancaldi, N. Leo, and P. Vavassori, *"Selective and Fast Plasmon-Assisted Photo-Heating of Nanomagnets"*, Nanoscale **11**, 7656 (2019).

7. C. D. Stanciu, F. Hansteen, A. V. Kimel, A. Kirilyuk, A. Tsukamoto, A. Itoh, and T. Rasing, *"All-Optical Magnetic Recording with Circularly Polarized Light"*, Phys. Rev. Lett. **99**, 047601 (2007).

8. A. Stupakiewicz, K. Szerenos, D. Afanasiev, A. Kirilyuk, and A. V. Kimel, *"Ultrafast Nonthermal Photo-Magnetic Recording in a Transparent Medium"*, Nature (London) **542**, 71 (2017).

9. D. Bossini, V. I. Belotelov, A. K. Zvezdin, A. N. Kalish, and A. V. Kimel, *"Magnetoplasmonics and Femtosecond Optomagnetism at the Nanoscale"*, ACS Photonics **3**, 1385 (2016).

10. N. Maccaferri, I. Zubritskaya, I. Razdolski, I. A. Chioar, V. Belotelov, V. Kapaklis, P. M. Oppeneer, and A. Dmitriev, *"Nanoscale Magnetophotonics"*, J. Appl. Phys. **127**, 080903 (2020).

11. A. Kirilyuk, A. V. Kimel, and T. Rasing, *"Ultrafast Optical Manipulation of Magnetic Order"*, Rev. Mod. Phys. **82**, 2731 (2010).

12. S. Mangin, M. Gottwald, C. H. Lambert, D. Steil, V. Uhlir, L. Pang, M. Hehn, S. Alebrand, M. Cinchetti, G. Malinowski, Y. Fainman, M. Aeschlimann, and E. E. Fullerton, *"Engineered Materials for All-Optical Helicity-Dependent Magnetic Switching"*, Nat. Mater. **13**, 287 (2014).

13. C. S. Davies, T. Janssen, J. H. Mentink, A. Tsukamoto, A. V. Kimel, A. F. G. van der Meer, A. Stupakiewicz, and A. Kirilyuk, *"Pathways for Single-Shot All-Optical Switching of Magnetization in Ferrimagnets"*, Phys. Rev. Appl. **13**, 024064 (2020).

14. L. Le Guyader, S. El Moussaoui, M. Buzzi, R. V. Chopdekar, L. J. Heyderman, A. Tsukamoto, A. Itoh, A. Kirilyuk, T. Rasing, A. V. Kimel, and F. Nolting, *"Demonstration of Laser Induced Magnetization Reversal in GdFeCo Nanostructures"*, Appl. Phys. Lett. **101**, 022410 (2012).

15. T. A. Ostler, J. Barker, R. F. L. Evans, R. W. Chantrell, U. Atxitia, O. Chubykalo-Fesenko, S. El Moussaoui, L. Le Guyader, E. Mengotti, L. J. Heyderman, F. Nolting, A. Tsukamoto, A. Itoh, D. Afanasiev, B. A. Ivanov, A. M. Kalashnikova, K. Vahaplar, J. Mentink, A. Kirilyuk, T. Rasing, and A. V. Kimel, *"Ultrafast Heating as a Sufficient Stimulus for Magnetization Reversal in a Ferrimagnet"*, Nat. Commun. **3**, 666 (2012).

16. L. Le Guyader, M. Savoini, S. El Moussaoui, M. Buzzi, A. Tsukamoto, A. Itoh, A. Kirilyuk, T. Rasing, A. V. Kimel, and F. Nolting, *"Nanoscale Sub-100 Picosecond All-Optical Magnetization Switching in GdFeCo Microstructures"*, Nat. Commun. **6**, 5839 (2015).

17. C. H. Lambert, S. Mangin, B. S. D. Ch. S. Varaprasad, Y. K. Takahashi, M. Hehn, M. Cinchetti, G. Malinowski, K. Hono, Y. Fainman, M. Aeschlimann, and E. E. Fullerton, *"All-Optical Control of Ferromagnetic Thin Films and Nanostructures"*, Science **345**, 1337 (2014).

18. R. John, M. Berritta, D. Hinzke, C. Muller, T. Santos, H. Ulrichs, P. Nieves, J. Walowski, R. Mondal, O. Chubykalo-Fesenko, J. McCord, P. M. Oppeneer, U. Nowak, and M. Münzenberg, *"Magnetisation Switching of FePt Nanoparticle Recording Medium by Femtosecond Laser Pulses"*, Sci. Rep. **7**, 4114 (2017).

19. M. Vomir, M. Albrecht, and J. Y. Bigot, *"Single Shot All Optical Switching of Intrinsic Micron Size Magnetic Domains of a Pt/Co/Pt Ferromagnetic Stack"*, Appl. Phys. Lett. **111**, 242404 (2017).

20. G. Baffou and H. Rigneault, *"Femtosecond-Pulsed Optical Heating of Gold Nanoparticles"*, Phys. Rev. B **84**, 035415 (2011).

21. E. Ringe, B. Sharma, A. I. Henry, L. D. Marks, and R. P. Van Duyne, *"Single Nanoparticle Plasmonics"*, Phys. Chem. Chem. Phys. **15**, 4110 (2013).

22. J. Y. Bigot, V. Halte, J. C. Merle, and A. Daunois, *"Electron Dynamics in Metallic Nanoparticles"*, Chem. Phys. **251**, 181 (2000).

23. L. H. F. Andrade, A. Laraoui, M. Vomir, D. Muller, J. P. Stoquert, C. Estournes, E. Beaurepaire, and J. Y. Bigot, *"Damped Precession of the Magnetization Vector of Superparamagnetic Nanoparticles Excited by Femtosecond Optical Pulses"*, Phys. Rev. Lett. **97**, 127401 (2006).

24. G. Baffou, P. Berto, E. B. Urena, R. Quidant, S. Monneret, J. Polleux, and H. Rigneault, *"Photoinduced Heating of Nanoparticle Arrays"*, ACS Nano **7**, 6478 (2013).

25. M. I. Tribelsky, A. E. Miroshnichenko, Y. S. Kivshar, B. S. Luk'yanchuk, and A. R. Khokhlov, *"Laser Pulse Heating of Spherical Metal Particles"*, Phys. Rev. X **1**, 021024 (2011).

26. This follows from $\Delta T = \Delta E/(mC_p) = \Delta E/(\rho V C_p) = (\Delta E/V)/(\rho C_p) = w/(\rho C_p)$, where $m$ is the mass of the sample, $C_p$ is the specific heat, and $\rho$ is the mass density of the considered material.

27. B. Koopmans, G. Malinowski, F. D. Longa, D. Steiauf, M. Faehnle, T. Roth, M. Cinchetti, and M. Aeschlimann, *"Explaining the Paradoxical Diversity of Ultrafast Laser-Induced Demagnetization"*, Nat. Mater. **9**, 259 (2010).

28. A. Kleibert, A. Balan, R. Yanes, P. M. Derlet, C. A. F. Vaz, M. Timm, A. Fraile Rodríguez, A. Béché, J. Verbeeck, R. S. Dhaka, M. Radovic, U. Nowak, and F. Nolting, *"Direct Observation of Enhanced Magnetism in Individual Size- and Shape-Selected 3d Transition Metal Nanoparticles"*, Phys. Rev. B **95**, 195404 (2017).

29. K. Ohta and H. Ishida, *"Matrix Formalism for Calculation of the Light-Beam Intensity in Stratified Multilayered Films, and Its Use in the Analysis of Emission-Spectra"*, Appl. Opt. **29**, 2466 (1990).

30. L. Le Guyader, A. Kleibert, F. Nolting, L. Joly, P. M. Derlet, R. V. Pisarev, A. Kirilyuk, T. Rasing, and A. V. Kimel, *"Dynamics of Laser-Induced Spin Reorientation in Co/Smfeo3 Heterostructure"*, Phys. Rev. B **87**, 054437 (2013).

31. T. Luo and G. Chen, *"Nanoscale Heat Transfer – From Computation to Experiment"*, Phys. Chem. Chem. Phys. **15**, 3389 (2013).

32. H. P. Myers and W. Sucksmith, *"The Spontaneous Magnetization of Cobalt"*, Proc. R. Soc. A **207**, 427 (1951).

33. C. A. F. Vaz, J. A. C. Bland, and G. Lauhoff, *"Magnetism in Ultrathin Film Structures"*, Rep. Prog. Phys. **71**, 056501 (2008).

34. W. Wernsdorfer, E. B. Orozco, K. Hasselbach, A. Benoit, B. Barbara, N. Demoncy, A. Loiseau, H. Pascard, and D. Mailly, *"Experimental Evidence of the Néel-Brown Model of Magnetization Reversal"*, Phys. Rev. Lett. **78**, 1791 (1997)

35. M. Iwaki, *"Estimation of the Atomic Density of Amorphous Carbon Using Ion Implantation, SIMS and RBS"*, Surf. Coat. Tech. **158-159**, 377 (2002).

36. C. Moelle, M. Werner, F. Szücs, D. Wittorf, M. Sellschopp, J. von Borany, H. J. Fecht, and C. Johnston, *"Specific Heat of Single-, Poly- and Nanocrystalline Diamond"*, Diam. Relat. Mater. **7**, 499 (1998).

37. E. D. Palik, in *Handbook of Optical Constants of Solids*, edited by E. D. Palik (Academic Press, Boston, 1998), pp. 313.

38. A. J. Bullen, K. E. O'Hara, D. G. Cahill, O. Monteiro, and A. von Keudell, *"Thermal Conductivity of Amorphous Carbon Thin Films"*, J. Appl. Phys. **88**, 6317 (2000).

39. W. H. Zhu, G. Zheng, S. Cao, and H. He, *"Thermal Conductivity of Amorphous Sio2 Thin Film: A Molecular Dynamics Study"*, Sci. Rep. **8**, 10537 (2018).

40. R. K. Endo, Y. Fujihara, and M. Susa, *"Calculation of the Density and Heat Capacity of Silicon by Molecular Dynamics Simulation"*, High Temp.-High Press. **35**, 505 (2003).

41. A. F. Rodríguez, A. Kleibert, J. Bansmann, and F. Nolting, *"Probing Single Magnetic Nanoparticles by Polarization-Dependent Soft X-Ray Absorption Spectromicroscopy"*, J. Phys. D: Appl. Phys. **43**, 474006 (2010).

42. A. Balan, P. M. Derlet, A. F. Rodríguez, J. Bansmann, R. Yanes, U. Nowak, A. Kleibert, and F. Nolting, *"Direct Observation of Magnetic Metastability in Individual Iron Nanoparticles"*, Phys. Rev. Lett. **112**, 107201 (2014).

43. C. A. F. Vaz, A. Balan, F. Nolting, and A. Kleibert, *"In Situ Magnetic and Electronic Investigation of the Early Stage Oxidation of Fe Nanoparticles Using X-Ray Photo-Emission Electron Microscopy"*, Phys. Chem. Chem. Phys. **16**, 26624 (2014).

44. L. Le Guyader, A. Kleibert, A. F. Rodríguez, S. El Moussaoui, A. Balan, M. Buzzi, J. Raabe, and F. Nolting, *"Studying Nanomagnets and Magnetic Heterostructures with X-Ray PEEM at the Swiss Light Source"*, J. Electron Spectrosc. **185**, 371 (2012).

45. A. Kleibert, J. Passig, K. H. Meiwes-Broer, M. Getzlaff, and J. Bansmann, *"Structure and Magnetic Moments of Mass-Filtered Deposited Nanoparticles"*, J. Appl. Phys. **101**, 114318 (2007).

46. H. Haberland, Z. Insepov, and M. Moseler, *"Molecular-Dynamics Simulation of Thin-Film Growth by Energetic Cluster Impact"*, Phys. Rev. B **51**, 11061 (1995).

47. V. N. Popok, I. Barke, E. E. B. Campbell, and K.-H. Meiwes-Broer, *"Cluster–Surface Interaction: From Soft Landing to Implantation"*, Surf. Sci. Rep. **66**, 347 (2011).

48. C. T. Chen, Y. U. Idzerda, H. J. Lin, N. V. Smith, G. Meigs, E. Chaban, G. H. Ho, E. Pellegrin, and F. Sette, *"Experimental Confirmation of the X-Ray Magnetic Circular-Dichroism Sum-Rules for Iron and Cobalt"*, Phys. Rev. Lett. **75**, 152 (1995).

49. J. Krempasky, U. Flechsig, T. Korhonen, D. Zimoch, C. Quitmann, and F. Nolting, *"Synchronized Monochromator and Insertion Device Energy Scans at SLS"*, AIP Conf. Proc. **1234**, 705 (2010).

50. A. A. Studna, D. E. Aspnes, L. T. Florez, B. J. Wilkens, J. P. Harbison, and R. E. Ryan, *"Low-Retardance Fused-Quartz Window for Real-Time Optical Applications in Ultrahigh-Vacuum"*, J. Vac. Sci.. Technol A **7**, 3291 (1989).

51. M. Vomir, M. Albrecht, and J.-Y. Bigot, *" Single Shot All Optical Switching of Intrinsic Micron Size Magnetic Domains of a Pt/Co/Pt Ferromagnetic Stack"*, Appl. Phys. Lett. **111**, 242404 (2017).

52. A. A. El-Gendy, M. Qian, Z.J. Huba, S. N. Khanna, and E. E. Carpenter, *" Enhanced Magnetic Anisotropy in Cobalt-Carbide Nanoparticles"*, Appl. Phys. Lett. **104**, 023111 (2014).

53. S. Rapp, M. Domke, M. Schmidt, and H. P. Huber, *"Physical Mechanisms During Fs Laser Ablation of Thin SiO$_2$ Films"*, Phys. Procedia **41**, 734 (2013).

54. T. J. Regan, H. Ohldag, C. Stamm, F. Nolting, J. Luning, J. Stohr, and R. L. White, *"Chemical Effects at Metal/Oxide Interfaces Studied by X-Ray-Absorption Spectroscopy"*, Phys. Rev. B **64**, 214422 (2001).

55. A. Bergmann, E. Martinez-Moreno, D. Teschner, P. Chernev, M. Gliech, J. F. de Araujo, T. Reier, H. Dau, and P. Strasser, *"Reversible Amorphization and the Catalytically Active State of Crystalline Co$_3$O$_4$ During Oxygen Evolution"*, Nat. Commun. **6**, 8625 (2015).

56. M. G. Khadra, *"Magnetic and Structural Properties of Size-Selected FeCo Nanoparticle Assemblies"*, PhD Thesi*s*, Universitè Claude Bernard Lyon, 2015.

57. M. Berritta, R. Mondal, K. Carva, and P. M. Oppeneer, *"Ab Initio Theory of Coherent Laser-Induced Magnetization in Metals"*, Phys. Rev. Lett. **117**, 137203 (2016).

58. P. W. Granitzka, E. Jal, L. Le Guyader, M. Savoini, D. J. Higley, T. M. Liu, Z. Chen, T. Chase, H. Ohldag, G. L. Dakovski, W. F. Schlotter, S. Carron, M. C. Hoffman, A. X. Gray, P. Shafer, E. Arenholz, O. Hellwig, V. Mehta, Y. K. Takahashi, J. Wang, E. E. Fullerton, J. Stohr, A. H. Reid, and H. A. Dürr, *"Magnetic Switching in Granular Fept Layers Promoted by near-Field Laser Enhancement"*, Nano Lett. **17**, 2426 (2017).

59. V. Papaefthimiou, T. Dintzer, V. Dupuis, A. Tamion, F. Tournus, A. Hillion, D. Teschner, M. Havecker, A. Knop-Gericke, R. Schlogl, and S. Zafeiratos, *"Nontrivial Redox Behavior of Nanosized Cobalt: New Insights from Ambient Pressure X-Ray Photoelectron and Absorption Spectroscopies"*, ACS Nano **5**, 2182 (2011).

60. A. Tamion, C. Raufast, M. Hillenkamp, E. Bonet, J. Jouanguy, B. Canut, E. Bernstein, O. Boisron, W. Wernsdorfer, and V. Dupuis, *"Magnetic Anisotropy of *Embedded Co Nanoparticles: Influence of the Surrounding Matrix"*, Phys. Rev. B **81**, 144403 (2010).

61. J. Hohlfeld, S. S. Wellershoff, J. Gudde, U. Conrad, V. Jahnke, and E. Matthias, *"Electron and Lattice Dynamics Following Optical Excitation of Metals"*, Chem. Phys. **251**, 237 (2000).

62. T. M. Liu, T. H. Wang, A. H. Reid, M. Savoini, X. F. Wu, B. Koene, P. Granitzka, C. E. Graves, D. J. Higley, Z. Chen, G. Razinskas, M. Hantschmann, A. Scherz, J. Stöhr, A. Tsukarnote, B. Hecht, A. V. Kimel, A. Kirilyuk, T. Rasing, and H. A. Dürr, *"Nanoscale Confinement of All-Optical Magnetic Switching in TbFeCo - Competition with Nanoscale Heterogeneity"*, Nano Lett. **15**, 6862 (2015).

63. O. Hovorka, S. Devos, Q. Coopman, W. J. Fan, C. J. Aas, R. F. L. Evans, X. Chen, G. Ju, and R. W. Chantrell, *"The Curie Temperature Distribution of FePt Granular Magnetic Recording Media"*, Appl. Phys. Lett. **101**, 052406 (2012).

64. J. Wang, W. Wu, F. Zhao, and G.-M. Zhao, *"Curie Temperature Reduction in $SiO_2$-Coated Ultrafine $Fe_3O_4$ Nanoparticles: Quantitative Agreement with a Finite-Size Scaling Law"*, Appl. Phys. Lett. **98**, 083107 (2011).

65. L. Le Guyader, S. El Moussaoui, M. Buzzi, M. Savoini, A. Tsukamoto, A. Itoh, A. Kirilyuk, T. Rasing, F. Nolting, and A. V. Kimel, *"Deterministic Character of All-Optical Magnetization Switching in GdFe-Based Ferrimagnetic Alloys"*, Phys. Rev. B **93**, 134402 (2016).